\documentstyle[aps,manuscript,epsfig]{revtex}
\tightenlines  
\begin{document}
\draft
\title{Search for the Flavor-Changing Neutral Current Decays 
$B^+ \rightarrow \mu^+\mu^-K^+$ and $B^0 \rightarrow \mu^+\mu^-K^{*0}$}
\author{\font\eightit=cmti8
\def\r#1{\ignorespaces $^{#1}$}
\hfilneg
\begin{sloppypar}
\noindent
T.~Affolder,\r {21} H.~Akimoto,\r {42}
A.~Akopian,\r {35} M.~G.~Albrow,\r {10} P.~Amaral,\r 7 S.~R.~Amendolia,\r {31} 
D.~Amidei,\r {24} J.~Antos,\r 1 
G.~Apollinari,\r {35} T.~Arisawa,\r {42}
% ...
\end{sloppypar}
\vskip .026in
\begin{center}
(CDF Collaboration)
\end{center}

\vskip .026in
\begin{center}
\r 1  {\eightit Institute of Physics, Academia Sinica, Taipei, Taiwan 11529, 
Republic of China} \\
\r 2  {\eightit Argonne National Laboratory, Argonne, Illinois 60439} \\
\r 3  {\eightit Istituto Nazionale di Fisica Nucleare, University of Bologna,
I-40127 Bologna, Italy} \\
\r 4  {\eightit Brandeis University, Waltham, Massachusetts 02254} \\
\r 5  {\eightit University of California at Los Angeles, Los 
Angeles, California  90024} \\  
\r 6  {\eightit Instituto de Fisica de Cantabria, University of Cantabria, 
39005 Santander, Spain} \\
\r 7  {\eightit Enrico Fermi Institute, University of Chicago, Chicago, 
Illinois 60637} \\
\r 8  {\eightit Joint Institute for Nuclear Research, RU-141980 Dubna, Russia}
\\
\r 9  {\eightit Duke University, Durham, North Carolina  27708} \\
\r {10}  {\eightit Fermi National Accelerator Laboratory, Batavia, Illinois 
60510} \\
\r {11} {\eightit University of Florida, Gainesville, Florida  32611} \\
\r {12} {\eightit Laboratori Nazionali di Frascati, Istituto Nazionale di Fisica
               Nucleare, I-00044 Frascati, Italy} \\
\r {13} {\eightit University of Geneva, CH-1211 Geneva 4, Switzerland} \\
\r {14} {\eightit Harvard University, Cambridge, Massachusetts 02138} \\
\r {15} {\eightit Hiroshima University, Higashi-Hiroshima 724, Japan} \\
\r {16} {\eightit University of Illinois, Urbana, Illinois 61801} \\
\r {17} {\eightit The Johns Hopkins University, Baltimore, Maryland 21218} \\
\r {18} {\eightit Institut f\"{u}r Experimentelle Kernphysik, 
Universit\"{a}t Karlsruhe, 76128 Karlsruhe, Germany} \\
\r {19} {\eightit Korean Hadron Collider Laboratory: Kyungpook National
University, Taegu 702-701; Seoul National University, Seoul 151-742; and
SungKyunKwan University, Suwon 440-746; Korea} \\
\r {20} {\eightit National Laboratory for High Energy Physics (KEK), Tsukuba, 
Ibaraki 305, Japan} \\
\r {21} {\eightit Ernest Orlando Lawrence Berkeley National Laboratory, 
Berkeley, California 94720} \\
\r {22} {\eightit Massachusetts Institute of Technology, Cambridge,
Massachusetts  02139} \\   
\r {23} {\eightit Institute of Particle Physics: McGill University, Montreal 
H3A 2T8; and University of Toronto,\\ Toronto M5S 1A7; Canada} \\
\r {24} {\eightit University of Michigan, Ann Arbor, Michigan 48109} \\
\r {25} {\eightit Michigan State University, East Lansing, Michigan  48824} \\
\r {26} {\eightit University of New Mexico, Albuquerque, New Mexico 87131} \\
\r {27} {\eightit The Ohio State University, Columbus, Ohio  43210} \\
\r {28} {\eightit Osaka City University, Osaka 588, Japan} \\
\r {29} {\eightit Universita di Padova, Istituto Nazionale di Fisica 
          Nucleare, Sezione di Padova, I-35131 Padova, Italy} \\
\r {30} {\eightit University of Pennsylvania, Philadelphia, 
        Pennsylvania 19104} \\   
\r {31} {\eightit Istituto Nazionale di Fisica Nucleare, University and Scuola
               Normale Superiore of Pisa, I-56100 Pisa, Italy} \\
\r {32} {\eightit University of Pittsburgh, Pittsburgh, Pennsylvania 15260} \\
\r {33} {\eightit Purdue University, West Lafayette, Indiana 47907} \\
\r {34} {\eightit University of Rochester, Rochester, New York 14627} \\
\r {35} {\eightit Rockefeller University, New York, New York 10021} \\
\r {36} {\eightit Rutgers University, Piscataway, New Jersey 08855} \\
\r {37} {\eightit Texas A\&M University, College Station, Texas 77843} \\
\r {38} {\eightit Texas Tech University, Lubbock, Texas 79409} \\
\r {39} {\eightit Istituto Nazionale di Fisica Nucleare, University of Trieste/
Udine, Italy} \\
\r {40} {\eightit University of Tsukuba, Tsukuba, Ibaraki 305, Japan} \\
\r {41} {\eightit Tufts University, Medford, Massachusetts 02155} \\
\r {42} {\eightit Waseda University, Tokyo 169, Japan} \\
\r {43} {\eightit University of Wisconsin, Madison, Wisconsin 53706} \\
\r {44} {\eightit Yale University, New Haven, Connecticut 06520} \\
\end{center}

}
\maketitle

\begin{abstract}
We report on a search for the flavor-changing neutral current decays
$B^+ \rightarrow \mu^+\mu^-K^+$ and $B^0 \rightarrow \mu^+\mu^-K^{*0}$
using $88$ ${\rm pb}^{-1}$ of data from $\bar{p}p$ collisions at
$\sqrt{s}=1.8$ TeV, collected with the Collider Detector at
Fermilab. Finding no evidence for these decays, we set upper limits on
the branching fractions $\mathcal{B}$$(B^+ \rightarrow
\mu^+\mu^-K^+)<5.2\times 10^{-6}$ and $\mathcal{B}$$(B^0 \rightarrow
\mu^+\mu^-K^{*0})<4.0\times 10^{-6}$ at the 90\% confidence level.
\end{abstract}


In the Standard Model of electroweak interactions, the flavor-changing
neutral current (FCNC) processes $B^+ \rightarrow \mu^+\mu^-K^+$ and $B^0
\rightarrow \mu^+\mu^-K^{*0}$ \cite{chargeconj}
are forbidden at tree level.  At higher
order, these decays occur through penguin and box diagrams with
predicted branching fractions 
$(0.3-0.5) \times 10^{-6}$ for $B^+ \rightarrow \mu^+\mu^-K^+$ and
$(1.0-1.5) \times 10^{-6}$ for $B^0 \rightarrow \mu^+\mu^-K^{*0}$ 
\cite{SM_predictions}.
Many deviations from Standard Model physics, such as charged Higgs
bosons or anomalous $W$ couplings, can increase these rates
significantly~\cite{SM_extensions}.  Because the spread in Standard
Model predictions, due mainly to uncertainties in hadronic form
factors, is well bounded, the observation of much larger branching
fractions would be a signal for physics outside the Standard Model.

To date, no observation of these decays has been reported.  The CLEO
collaboration~\cite{CLEO_rareB} has set the limits 
${\cal B}(B^+ \rightarrow \mu^+\mu^-K^+) < 0.97 \times 10^{-5}$ and 
${\cal B}(B^0 \rightarrow \mu^+\mu^-K^{*0}) < 0.95 \times 10^{-5}$ 
at the 90\%
confidence level (C.L.).  The CDF collaboration~\cite{CDF_prl} has previously
searched for these decays in an $18$ ${\rm pb}^{-1}$ data sample recorded in
1992-1993 and set the 90\% C.L. limits 
$ {\cal B}(B^+ \rightarrow
\mu^+\mu^-K^+) < 1.0 \times 10^{-5}$ and $ {\cal B}(B^0 \rightarrow
\mu^+\mu^-K^{*0}) < 2.5 \times 10^{-5}$.  The result presented in this
Letter is based on an $88$ ${\rm pb}^{-1}$ combined data sample from
the 1992-1993 and 1994-1995 running periods and supersedes the earlier
CDF result.

Briefly, the method used in this search is as follows.  The CDF
trigger selects events containing two muon candidates, which we
combine offline with $K^{+}$ and $K^{*0}$ candidates to form $B^{+}$
and $B^{0}$ candidates.  Using the topologically
identical resonant processes $B^+ \rightarrow J/\psi K^+ \rightarrow
(\mu^+\mu^-)K^+$ and $B^{0} \rightarrow J/\psi K^{*0} \rightarrow
(\mu^+\mu^-)K^{*0}$ for normalization, we measure ratios of branching
fractions, thus canceling most uncertainties due to $B$ meson
production and detection efficiency.  Small acceptance and trigger
efficiency differences due to the different decay kinematics are
corrected with a Monte Carlo calculation.

 The CDF detector has been described in detail elsewhere~\cite{CDF}.
Detector components most relevant to this search are the muon chambers
and charged particle tracking system.  The muon system is composed of
drift chambers outside the hadron calorimeter, allowing the
reconstruction of track segments for penetrating particles in the
pseudorapidity~\cite{coordinates} range $|\eta|<1$.  The minimum
transverse momentum ($p_T$) of detectable muons varies between $1.4$
and $2.2$ GeV$/c$ as a function of $\eta$, depending on the amount of
material between the interaction point and the muon chambers.  Charged
particles are reconstructed in the central tracking chamber (CTC) and
the silicon vertex detector (SVX)\cite{SVX}, both located inside a $1.4$~T
solenoidal magnetic field.  The CTC, covering $|\eta|<1.1$, is a
cylindrical drift chamber containing 84 layers grouped into nine
alternating superlayers of axial and stereo wires.  The SVX consists
of four layers of single-sided silicon microstrip detectors mounted at
radii between 2.9 and 7.9 cm from the beam. It provides track
measurements in the \mbox{$r$-$\phi$} plane with impact parameter resolution
(13 + 40/$p_T$) $\mu$m, where $p_T$ is in GeV$/c$.
 The impact parameter of a track is defined as the distance of 
closest approach to the beam in a plane transverse to 
the beam.

CDF uses a three-level trigger system. 
The muon pair trigger requires the presence of two track segments in the
muon chambers at the first level and matching CTC track candidates
found by the central fast track processor (CFT)\cite{CFT}
for one or both muon segments at the second level.
 The third-level trigger tightens the matching criteria using fully
reconstructed CTC tracks.

Muons inside the acceptance of the muon chambers are found
by the first-level trigger with an efficiency that rises from 
about $40\%$ at $p_T = 1.5$ GeV$/c$ to a plateau value of $93\%$
for muon transverse momenta exceeding 3 GeV$/c$.
The efficiency for finding the corresponding tracks with the CFT
at level two 
rises from about $50\%$ at $p_T = 2$ GeV$/c$ to 
about $96\%$ for $p_T > 2.3$ GeV$/c$. 
If one muon does not have a matching CFT track
a higher $p_T$ threshold is imposed for the other muon,
leading to an efficiency that rises from about $50\%$ at 
$p_T = 2.6$~GeV$/c$ and reaches the same plateau value for $p_T>3.1$~GeV$/c$.
The first type of trigger, requiring two CFT tracks, was not
implemented during the 1992-1993 running period.
A muon pair is called resonant when its invariant mass is close to the
$c\bar{c}-$resonances $J/\psi$ or $\psi'$.
Event selection is identical for resonant and non-resonant muon pairs.

To reconstruct $\mu^+ \mu^- K^+$ candidates, we combine
the two muon candidates selected by the trigger with an additional
charged track, assumed to be a kaon.  The kaon track must satisfy $p_T
> 2$ GeV$/c$, and the $B$ candidate must satisfy $p_T > 6$ GeV/$c$.

The reconstruction of the $K^{*0}$ in the 
$B^0 \rightarrow\mu^+\mu^-K^{*0}$ process is via the
$K^{*0} \rightarrow K^+ \pi^-$ decay.
We require $p_T > 0.5$\ GeV/$c$ for each track, 
and to reduce random combinations,
we require $p_T > 2$\ GeV/$c$ for the $K^{*0}$ candidate.  
To allow for the $K^*(892)^0$ width, we require the
$K^{*0}$ candidate mass to be between 796 and 996 MeV/$c^2$.  
No particle identification is used, and each pair of tracks is
considered as $K^+\pi^-$ and $\pi^+K^-$, respectively.
We choose the $K^{*0}/\bar{K}^{*0}$-daughter mass assignment 
with invariant mass
closer to the world-average $K^{*0}$ mass.

An isolation requirement is a well
established means of improving the signal to background ratio in
reconstructing B decays.
We require isolation $I>0.6$, where $I$ is the transverse momentum of the $B$
candidate divided by the scalar sum of the transverse momenta of the
$B$ and all other tracks in a cone
$\sqrt{(\Delta\eta)^2+(\Delta\phi)^2}<1.0$ around the $B$ momentum.
The measured efficiency from our data for the requirement $I>0.6$ 
is $92\pm 6$\% for $B$ mesons with $p_T > 6 $\ GeV/$c$
and $\left| \eta \right| < 1$.

A possible background for the rare decay search comes from
$B\rightarrow J/\psi K$ decays in which the kaon is misidentified as a
muon and the muon is taken as the kaon instead.  We apply a cut to 
remove candidates that are compatible with this hypothesis. 
In fact, no non-resonant candidate showed this ambiguity after all
other selection cuts were applied.

We require $B$ daughters to be consistent with originating at a displaced 
decay vertex and inconsistent with originating at the beam line.
All tracks forming the $B$ candidate must include hits on at least
three SVX layers, to ensure a good impact parameter measurement, and
must miss the beam line with at least $2\sigma$ signficance, where
$\sigma$ is the estimated impact parameter resolution. The
tracks are fit to a common decay vertex.  The fit assumes that the $B$
meson is produced at the beam line and constrains the $B$ momentum to
be parallel to its line of flight in the \mbox{$r$-$\phi$} plane.  The fit
quality must be acceptable ($\chi^2 < 20$ with 4 degrees of freedom
for $\mu^+\mu^- K^+$ and 6 for $\mu^+\mu^- K^{*0}$)
and the transverse decay length must
exceed 400 $\mu$m.  
The transverse decay length is the distance between the beam line 
and the $B$ decay vertex measured in the $r-\phi$ plane.
The beam spot of the Tevatron has a transverse
size of approximately 25 $\mu$m (R.M.S.) and a mean that is measured with
negligible uncertainty~\cite{CDF_lifetime_PRD}.  The impact-parameter
and decay-length criteria are chosen to minimize the expected upper
limit using the signal efficiency from a Monte Carlo calculation and a
background estimate from the $B$ sideband mass region $5.38$--$5.88$
GeV/$c^2$.

\begin{figure}
\epsfig{file=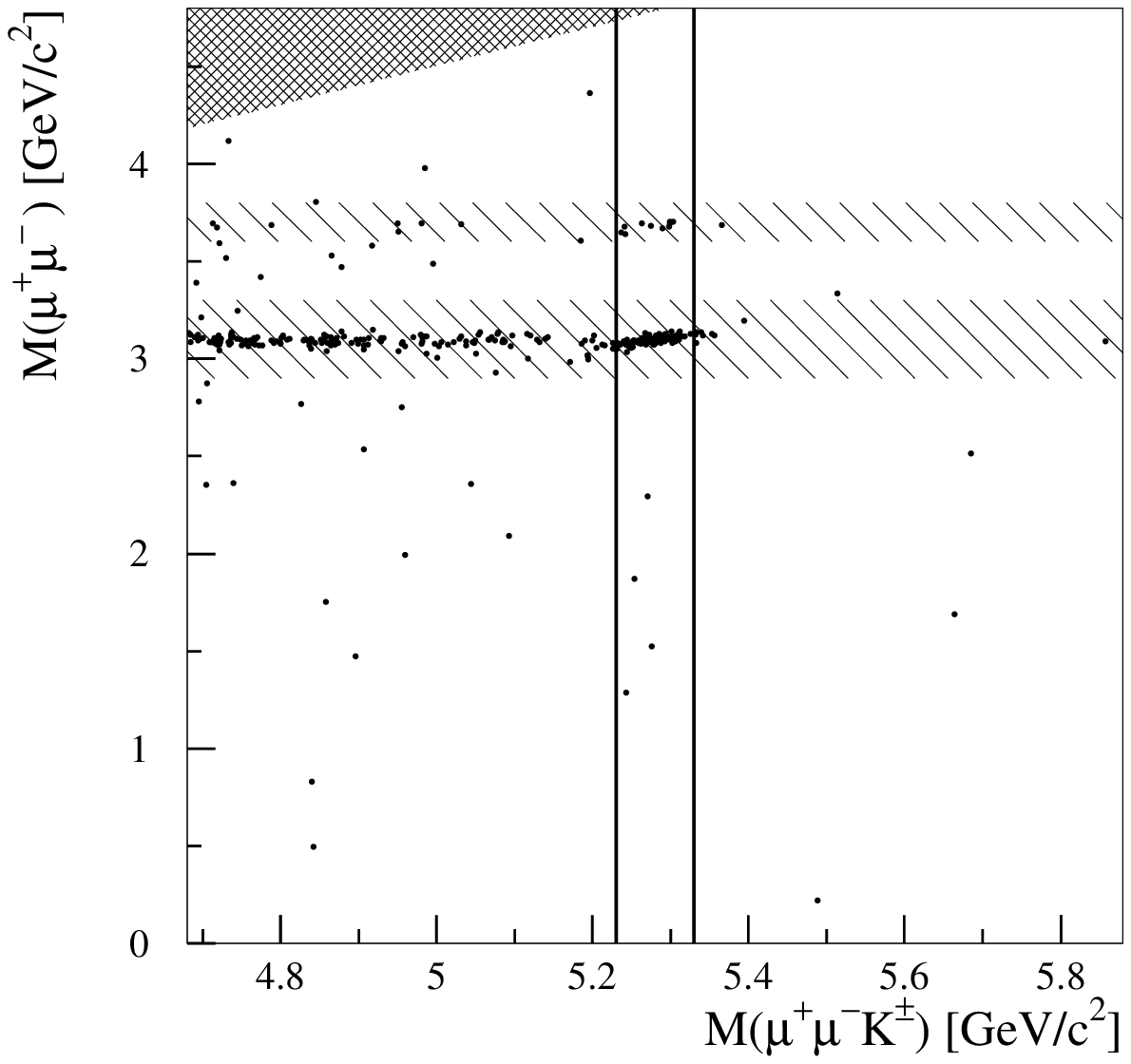,width=15cm}
\caption{Scatter plot showing the $B^{\pm}\rightarrow \mu^+\mu^- K^{\pm}$
candidates. The hatched horizontal bands are the excluded regions
around the $J/\psi$ and $\psi'$ resonances. The signal region is
between the vertical lines.  
The cross-hatched area in the upper left corner is kinematically forbidden.
Entries with $M(\mu^+\mu^-K^{\pm}) < 5.23$ GeV$/c^2$ and
dimuon masses in the $J/\psi$ region can be attributed to incompletely
reconstructed $B\rightarrow J/\psi X$ decays. }
\label{k_1b_ip_2d}
\end{figure}

\begin{figure}
\epsfig{file=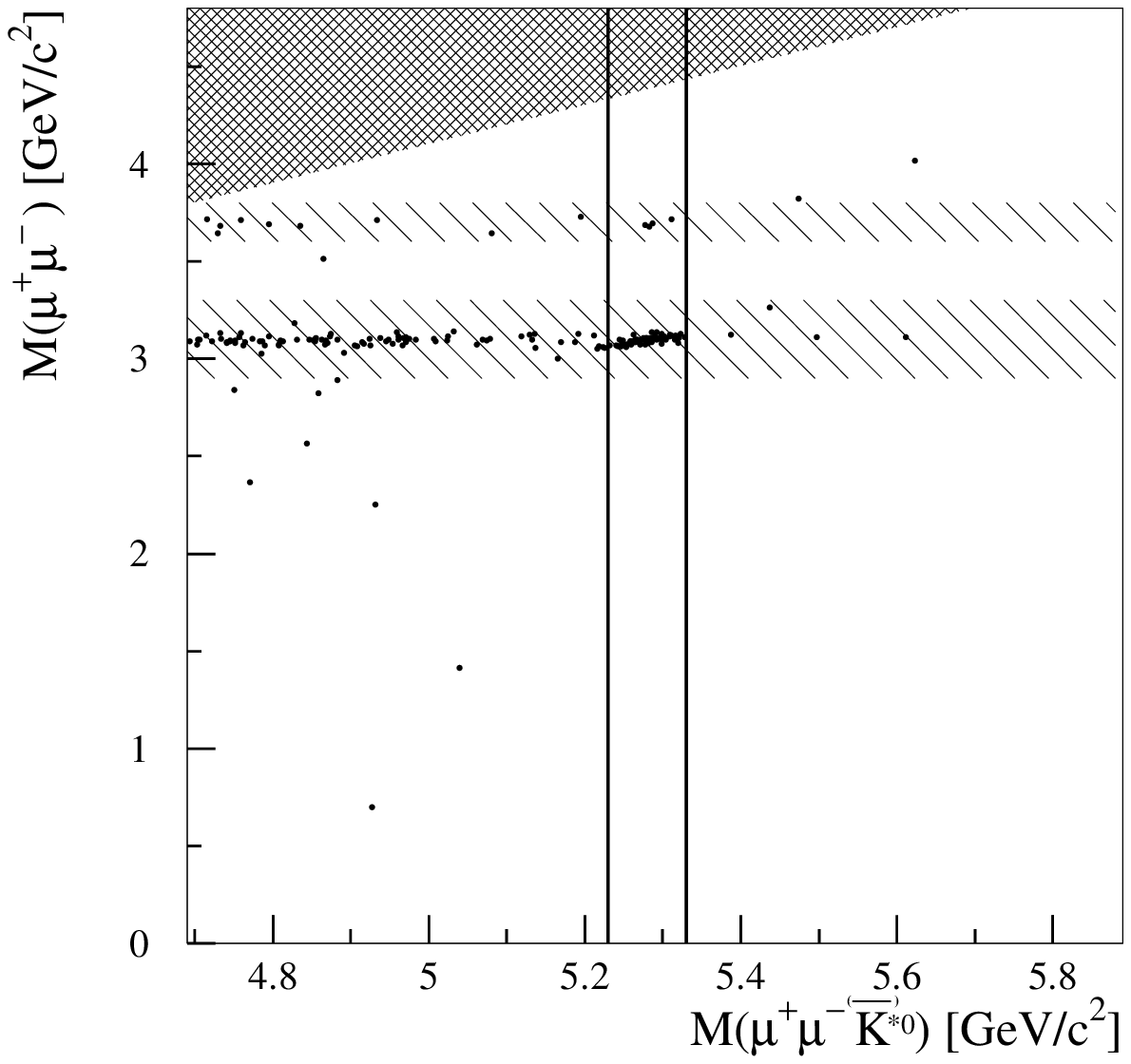,width=15cm}
\caption{Scatter plot showing the $B^0 \rightarrow \mu^+ \mu^- K^{*0}$
and $\bar{B}^0 \rightarrow \mu^+ \mu^- \bar{K}^{*0}$ candidates. The hatched
horizontal bands are the excluded regions around the $J/\psi$ and $\psi'$ 
resonances.  
}
\label{kstar_1b_ip_2d}
\end{figure}

 Figures \ref{k_1b_ip_2d}  and  \ref{kstar_1b_ip_2d} show scatter plots
of the $\mu^+\mu^- K^+$ and $\mu^+\mu^- K^{*0}$ mass versus the
$\mu^+\mu^-$ mass for candidates passing all requirements.
 The signal region for B mesons, whose half width is chosen
to be twice the invariant mass resolution, 
includes reconstructed $B$ meson masses
between $5.23$ and $5.33$ GeV/$c^2$.

The decays $B^+\rightarrow J/\psi(\psi') K^+$ 
and $B^0 \rightarrow J/\psi(\psi')K^{*0}$ 
followed by $J/\psi(\psi') \rightarrow \mu^+ \mu^-$ 
have product branching
fractions two orders of magnitude larger than the Standard Model
predictions for the 
non-resonant $B^+ \rightarrow \mu^+\mu^-K^+$ and 
$B^0 \rightarrow \mu^+\mu^-K^{*0}$ decays.
Candidates with an invariant mass of the muon pair
in the regions $2.9$--$3.3~{\rm GeV}/c^2$ or $3.6$--$3.8~{\rm
GeV}/c^2$ are therefore excluded from the rare decay search.  Events
that have dimuon masses within 100 MeV$/c^2$ of the $J/\psi$ mass are
used as the normalization sample.

In the $B$ mass region for $\mu^{+}\mu^{-} K^{+}$ final states, we count
4 non-resonant candidates and count 122 $J/\psi$ candidates. 
In case of the $\mu^+\mu^-K^{*0}$ final state, 
no non-resonant candidate is found in the $B$-mass region,
while we count 76 $J/\psi$ candidates.
The background for the resonant decays is estimated by counting events
in the $B$ mass sideband $5.38-5.88$~GeV/$c^2$ and 
scaling for the relative sizes of the $B$ signal region and sideband
regions. The estimates obtained this way are $0.4$ events for 
$B^+ \rightarrow J/\psi K^+$ and $0.6$ events for 
$B^0 \rightarrow J/\psi K^{*0}$.

The excess of resonant events in the sideband below the $B$ signal region
over the sideband above the $B$-mass is understood
to come from incompletely reconstructed $B \rightarrow J/\psi X$
decays. The non-resonant excess below the $B$ signal region is not
understood. The high mass sideband, five times as wide as the 
signal region, contains 4 non-resonant $\mu^+\mu^- K^+$ and 
2 non-resonant $\mu^+ \mu^- K^{*0}$ candidates.
Note that simply extrapolating these sideband rates may underestimate the 
background in the signal region, because the high-mass sideband 
is used to optimize the selection criteria.

Assuming all observed candidates in the signal region are signal events,
we set upper limits on the numbers of observable FCNC decays 
using Poisson statistics
and conclude that $\bar{N}(\mu^+\mu^- K^+) < 8.0$ and 
$\bar{N}(\mu^+\mu^- K^{*0})< 2.3$ 
at the 90\% confidence level.

We use a Monte Carlo simulation to calculate $R_\epsilon$, the ratio
of the total trigger and reconstruction efficiency for the non-resonant
mode to that of the $J/\psi$ mode, including the effect of the
restricted dimuon mass range in the case of non-resonant decays.  
In the Monte Carlo simulation, $B$ decays are sampled from a phase 
space distribution and weighted
according to the decay amplitude, using form factors and Wilson
coefficients taken from Ref.~\cite{GreubWyler}.  We find
$R_\epsilon(\mu^+\mu^-K^+) = 0.79$ and $R_\epsilon(\mu^+\mu^-K^{*0}) =
0.65$.
For observable numbers $\bar{N}(\mu^+ \mu^- K^{(*)})$ and $\bar{N}(J/ \psi K^{(*)})$
of non-resonant and $J/\psi$ decays, one obtains the branching
fractions for the FCNC decays,
\begin{displaymath}
{\cal B}(B\rightarrow \mu^+\mu^- K^{(*)}) =  
  \frac{\bar{N}(\mu^+ \mu^- K^{(*)})}{ N(J/ \psi K^{(*)}) 
  \times R_\epsilon } \times {\cal B}(B\rightarrow J/\psi K^{(*)}) 
  \times {\cal B}(J/\psi \rightarrow \mu^+ \mu^-).
\end{displaymath}

Whereas the central values of $R_\epsilon$ assume Standard Model
couplings, distributions of kinematic variables in the decays
$B\rightarrow \mu^+\mu^-K^{(*)}$ can be sensitive to non-standard
short-distance interactions~\cite{GreubWyler,Burdman_95}.  This
affects the trigger efficiency, the acceptance of the event selection,
and the extrapolation over excluded dimuon mass regions. To study such
model dependence, we varied the signs of the relevant Wilson
coefficients at $M_W$, as suggested in Ref.~\cite{GreubWyler}, and
found the change in efficiency to be much less than the change in the
predicted branching fraction.  No variation in $R_\epsilon$ larger
than 4\% (12\%) was found for $\mu^+\mu^- K^+$ $(\mu^+\mu^- K^{*0})$.

While this
procedure does not cover the most general set of models, we conclude
that the limits reported in this Letter, which are determined assuming
Standard Model couplings, will change little when applied to many
extended models.  We do not treat model dependence as a source of
error.

The impact of hadronic form factor uncertainties on $R_\epsilon$ was studied by
replacing the form factors from Ref.~\cite{GreubWyler} by the two sets
given in Ref.~\cite{Melikhov}.  We use the observed change in
$R_\epsilon$, less than 4\% (8\%) for the $K$ ($K^*$) mode, as an
estimate of the systematic error due to form factors.

Dominant systematic errors are the 11\% (14\%) uncertainty on the
product of world average values for the branching fractions
$B\rightarrow J/\psi K^+$ $(K^{*0})$ and $J/\psi \rightarrow \mu^+
\mu^-$ \cite{PDG}, and the statistical uncertainty on the number of
resonant events.  All other sources of error are found to make much smaller
contributions.  We estimate the total systematic error, including
hadronic uncertainties, to be 15\% (20\%) for $\mu^+\mu^- K(K^*)$, 
and include it in the
determination of an upper limit on the branching fraction, using the
method of Ref.~\cite{Cousins}.

Assuming all observed candidates to be signal events we find
\begin{displaymath}
  {\cal B}(B^+ \rightarrow \mu^+\mu^-K^+) < 5.2 \times 10 ^{-6}
\end{displaymath}
and
\begin{displaymath}
  {\cal B}(B^0 \rightarrow \mu^+\mu^-K^{*0}) < 4.0 \times 10 ^{-6}
\end{displaymath}
at 90\% confidence level.
These are the strictest limits on these decay branching fractions
to date.  Assuming Standard Model predictions, 
the expectation is to observe approximately $0.5$ signal events 
in each channel.  

We thank the Fermilab staff and the technical staffs of the
participating institutions for their vital contributions.  This work was
supported by the U.S. Department of Energy and National Science Foundation,
the Italian Instituto Nazionale di Fisica Nucleare; the Ministry of Education,
Science and Culture of Japan, the Natural Sciences and Engineering Research
Council of Canada, the National Science Council of the Republic of
China, the A. P. Sloan Foundation, and the Alexander von Humboldt-Stiftung.


%
%
%

%
%


\begin{references}

\bibitem{chargeconj}
Throughout this paper, reference to a decay mode implies the charge
conjugate process as well.

\bibitem{SM_predictions}
A.~Ali, Nucl. Phys. B, Proc. Suppl. {\bf 59}, 86 (1997);
%
D.~Melikhov, N.~Nikitin and S.~Simula, Phys. Rev. D{\bf 57}, 6814 (1998);
%
T.~M.~Aliev, M.~Savci and A.~\"Ozpineci, Phys. Rev. D{\bf 56}, 4260 (1996);
%
P.~Colangelo, F.~De~Fazio, P.~Santorelli and E.~Scrimieri, Phys. Rev. D{\bf 53}, 3672 (1996);
%
C.~Q.~Geng and C.~P.~Kao, Phys. Rev. D{\bf 54}, 5636 (1996).

\bibitem{SM_extensions}
G.~Burdman, Phys. Rev. D{\bf 59}, 035001 (1999); 
T.~M.~Aliev, M.~Savci, A.~\"Ozpineci and I.~L.~Koru, Phys. Lett. B{\bf
410},
216 (1997);  
N.~G.~Deshpande, K.~Panose and J.~Trampetic, Phys. Lett. B{\bf 308}, 322
(1993); 
Y.~Okada, Y.~Shimizu and M.~Tanaka, Phys. Lett. B{\bf 405}, 297 (1997).


\bibitem{CLEO_rareB}
R.~Godang {\it et al.}, CLEO CONF 98-22; to be published in 
{\it Proceedings of the XXVII International Conference on High Energy 
Physics}, Vancouver, Canada, 1998, edited by 
 A.~Astbury, D.~Axen and J.~Robinson (World Scientific, 1999).
\bibitem{CDF_prl}
F.~Abe, {\it et al.}, Phys. Rev. Lett. {\bf 76}, 4675 (1996).


\bibitem{CDF}
F.~Abe {\it et al.}, Nucl. Instrum. Meth. A{\bf 271}, 387 (1988);
F.~Abe {\it et al.}, Phys. Rev. D{\bf 50}, 2966 (1994).


\bibitem{coordinates} In CDF, the positive $z$ axis lies along the
proton direction, $r$ is the radius from this axis, $\theta$ is the
polar angle, and $\phi$ is the azimuthal angle.  The pseudorapidity,
$\eta$, is defined as $\eta = \ln\cot(\theta/2)$.  Directions perpendicular
to the $z$ axis are called transverse.

\bibitem{SVX}
D.~Amidei {\it et al.}, Nucl. Instrum. Meth. A{\bf 350}, 73 (1994).

\bibitem{CFT}
 G.W.~Foster, J.~Freeman, C.~Newman-Holmes, and J.~Patrick, 
Nucl. Instr. Meth. A{\bf 269}, 93 (1988).

\bibitem{CDF_lifetime_PRD}
F.~Abe {\it et al.}, Phys. Rev. D{\bf 57}, 5382 (1998).

\bibitem{GreubWyler}
C.~Greub, A.~Ioannissian and D.~Wyler, Phys. Lett. B{\bf 346}, 149 (1995).

\bibitem{Burdman_95}
G.~Burdman, Phys. Rev. D{\bf 52}, 6400 (1995).

\bibitem{Melikhov}
D. Melikhov, N. Nikitin and S. Simula, Phys. Rev. D{\bf 57}, 6814 (1998).

\bibitem{PDG}
C.~Caso {\it et al.}, Eur. Jour. Phys C{\bf 3}, 1 (1998).

\bibitem{Cousins} R.~D.~Cousins and V.~L.~Highland, Nucl. Instrum. Meth. A{\bf 320}, 331 (1992).

\end{references}
\end{document}